\documentclass[aps,reprint,groupedaddress,floatfix,showpacs]{revtex4-1}
\usepackage{graphicx,float}
\usepackage{epsfig,amsmath,booktabs}
\usepackage{times,xcolor,hyperref}

\usepackage{amsmath}
\usepackage{SIunits}

\begin{document}

\title[]
{Clustered impurities and carrier transport in supported graphene}

\author{N. Sule}
\author{S. C. Hagness}
\author{I. Knezevic}\email[Email: ]{knezevic@engr.wisc.edu}
\affiliation{Department of Electrical and Computer Engineering, University of Wisconsin--Madison, Madison, WI 53706-1691, USA}

\begin{abstract}
We investigate the effects of charged impurity distributions and carrier-carrier interactions on electronic transport in graphene on SiO$_2$ by employing a self-consistent coupled simulation of carrier transport and electrodynamics. We show that impurity clusters of characteristic width 40--50 nm generate electron--hole puddles of experimentally observed sizes. In the conductivity versus carrier density dependence, the residual conductivity and the linear-region slope are determined by the impurity distribution, and the measured slope can be used to estimate the impurity density in experiment. Furthermore, we show that the high-density sublinearity in the conductivity stems from carrier-carrier interactions.
\end{abstract}
\date{\today}

\maketitle
\section{Introduction}

Graphene, a single sheet of carbon with a honeycomb lattice, is a two-dimensional (2D) material whose high carrier mobility and carrier density tunable by a back gate  \cite{RevModPhys.81.109,RevModPhys.83.407,doi:10.1021/nl102824h,doi:10.1146/annurev-conmatphys-062910-140458,Schwierz:2010ys} make it attractive for electronic device application    \cite{Bonaccorso:2010kx,Kim:2009vn,doi:10.1021/nl101559n,Lin05022010,Sensale-Rodriguez:2012ly,0022-3727-45-30-303001,Schedin:2007zr,Pumera2011308}. Large-area, good-quality graphene is commonly fabricated by chemical vapor deposition (CVD) on metal substrates \cite{yu:113103,Kim:2009vn,Li05062009}, followed by transfer onto insulating substrates using polymers, such as poly-dimethyl siloxane (PDMS) or poly-methyl methacrylate (PMMA). An important concern with these processing methods is the contamination of graphene with organic molecules \cite{meyer2007structure}, residues of the transfer polymer and metal ions \cite{doi:10.1021/nl203733r}, or charged impurities trapped in the supporting substrate \cite{casiraghi2007raman}.

Impurities near graphene are believed to be responsible for several observed  transport properties. Spatial inhomogeneities in the carrier density, known as electron--hole puddles, are formed due to the presence of charged impurities in the substrate  \cite{Martin:2008ve,Zhang:2009qf,PhysRevLett.101.166803}. The  charged impurities and the resulting electron--hole puddles have been linked to the observed non-universal minimum conductivity (also known as residual conductivity) of graphene close to the Dirac point \cite{Adam20112007}. However, high-resolution scanning tunneling microscopy (STM) studies \cite{PhysRevB.83.155409} have shown that electron--hole puddles near the Dirac point are typically $20\usk\nano\meter$ in diameter, while theoretical calculations using a random charged impurity distribution near graphene result in electron--hole puddle sizes of only about $9\usk\nano\meter$ \cite{PhysRevLett.101.166803}. This evidence suggests that the underlying charged impurities may be clustered. It has also been shown that PMMA and metal ion residue can persist on graphene samples even post-annealing \cite{doi:10.1021/nl203733r} and transmission electron microscopy (TEM) images \cite{doi:10.1021/nl203733r} show that the residue is not uniformly distributed, but forms clusters. Furthermore, the formation of gold clusters has been shown to affect the electron mobility in graphene \cite{mccreary2010effect}.

The linear dependence of conductivity, $\sigma$, on carrier density, $n$, has been attributed to carrier scattering with charged impurities \cite{tsuneya2006screening,PhysRevLett.98.186806}. However, experimental measurements distinctly display a sublinear $\sigma(n)$ dependence away from the charge-neutrality point \cite{Chen:2008fk,2010NatNa...5..722D,tan2007measurement}. The origin of the  sublinear $\sigma(n)$ behavior is still under debate: it has been ascribed to different physical mechanisms, such as electron scattering with residual  organic molecules \cite{wehling2010resonant} or the effect of spatial correlations in the distribution of the charged impurities near graphene \cite{PhysRevLett.107.156601,radchenko2012influence}.

In this paper, we employ numerical simulation of coupled carrier transport and electrodynamics to investigate the role of carrier-carrier and carrier-ion Coulomb interactions on the room-temperature, low-field transport in graphene on SiO$_2$, with focus on the effect of impurity clustering. We solve the Boltzmann equation for carrier transport by using the ensemble Monte Carlo (EMC) method, coupled with the electrodynamics solver that incorporates the finite-difference time-domain (FDTD) solution to Maxwell's curl equations and molecular dynamics (MD) for short-range carrier-carrier and carrier-ion interaction. We show that clustered distributions of impurities with an average cluster size of 40--50 nm result in the formation of $20\usk\nano\meter$-wide electron--hole puddles, the size observed in several experiments \cite{PhysRevB.83.155409,Xue:2011uq,Zhang:2009qf}. We demonstrate that the sublinear behavior of conductivity at high carrier densities, which becomes more pronounced with decreasing impurity density \cite{Chen:2008fk,2010NatNa...5..722D}, stems from short-range carrier-carrier interactions. Also, we show that the linear portion of the conductivity versus carrier density curve is governed by carrier-ion interactions, with the slope and the residual conductivity dependent on both the sheet impurity density and the impurity distribution. We characterize the dependence of the conductivity slope on the impurity density for uniform random and clustered distributions, which can be used to estimate the impurity density in experiment.

This paper is organized as follows: In Sec. \ref{sec:simulation framework}, we overview the EMC, FDTD, and MD techniques and their coupling (Sec. \ref{sec:emdfdtdmd}), and describe the generation of a clustered impurity distribution (Sec. \ref{sec:clustered imps}). In Sec. \ref{sec:results}, we discuss electron--hole puddle formation (Sec. \ref{sec:e-h puddles}), the role of impurity clustering in low-carrier-density transport (Sec. \ref{sec:imp distribution}, sublinearity in conductivity and its connection to the short-range carrier-carrier interaction (Sec. \ref{sec:sublinearity}, and how to estimate impurity density from the linear-region conductivity slope (Sec. \ref{sec:inverse slope}). We conclude with Sec. \ref{sec:conclusion}.

\section{THE SIMULATION FRAMEWORK}\label{sec:simulation framework}

Our goal is to accurately simulate room-temperature electron and hole transport in supported graphene with charged impurities in the substrate, with focus on impurity clustering and Coulomb interactions (carrier--ion and carrier--carrier). Experiments have shown that charged impurities are the dominant source of disorder in supported graphene \cite{tan2007measurement,Chen:2008fk,JangPRL2008}. As shown by Kohn and Luttinger \cite{KohnLuttingerPR1957}, the Boltzmann transport equation can be derived quite generally from the density-matrix formalism for electrons in the presence of dilute uncorrelated charged impurities. Indeed, at moderate carrier densities in graphene, transport is diffusive and well-described by the Boltzmann transport equation, with the conductivity being linear in the carrier density owing to carrier--ion interactions \cite{tsuneya2006screening,RevModPhys.83.407}. In the vicinity of the Dirac point, the average carrier density can be considerably lower than the impurity density and charge inhomogeneities referred to as puddles govern transport. However, the effective medium theory \cite{PhysRevLett.101.166803,RossiPRB2009,RevModPhys.83.407,Adam20091072} argues that, while the average carrier density for the entire sample may be low, carrier density within an individual puddle is fairly uniform and on the order of the impurity sheet density, and the Boltzmann transport picture remains applicable \cite{Adam20112007}.

Therefore, we will assume the diffusive transport regime, captured through the Boltzmann transport equation, throughout the range of carrier and impurity densities and distributions  considered here. In fact, we find that clustered impurities result in sizeable puddles with the carrier density that is nearly uniform and is of order the impurity density, in agreement with the effective medium theory. The assumption of diffusive transport is further strengthened by the fact that we are at room temperature and working with macroscopic samples with size greater than the mean free path \cite{Adam20112007}.

\subsection{EMC/FDTD/MD for Graphene on SiO$_2$}\label{sec:emdfdtdmd}

In order to simulate diffusive carrier transport and electrodynamics in supported graphene, we employ a coupled EMC/FDTD/MD technique \cite{willis:063714,emcfdtdmdgraphenejcel}. In a nutshell, EMC solves the Boltzmann transport equation, FDTD solves Maxwell's curl equations, while MD accounts for the interaction of charges when very close to one another. The coupled EMC/FDTD/MD technique was successfully used to calculate the high-frequency conductivity of bulk silicon, with very good agreement to experimental data \cite{willis:063714,Willis:2013uq}. Below, we briefly describe the key elements of the constituent techniques and refer the interested reader to references \cite{willis:063714} and \cite{emcfdtdmdgraphenejcel} for extensive computational detail.

EMC is a stochastic numerical technique widely used for solving the Boltzmann transport equation \cite{RevModPhys.55.645}. In EMC, a large ensemble (typically of order $10^5$) of carriers is tracked over time as they experience periods of free flight interrupted by scattering events.     Free-flight duration, the choice of the relaxation mechanisms, and carrier momentum direction  post-scattering are sampled stochastically according to appropriate distributions. During free flight, carriers interact with the local electromagnetic fields via the Lorentz force, $\vec{F}=q(\vec{E}+\vec{v}\times\vec{B})$, where $q$, $\vec{v}$, $\vec{E}$, and $\vec{B}$ are the carrier charge, carrier velocity, electric field, and magnetic flux density, respectively. The fields are calculated using the electrodynamics solver that includes the FDTD and MD components. The evolution of physical properties of interest, such as the carrier average drift velocity or kinetic energy, are calculated by averaging over the ensemble.

The FDTD method \cite{taflove2005computational} is a popular and highly accurate grid-based technique for solving Maxwell's curl equations. In FDTD, Maxwell's equations are discretized in both time and space by centered differences using the fully explicit Yee algorithm \cite{1138693}:  the components of electric and magnetic fields, $\vec E$ and $\vec H$, are spatially staggered and solved for in time using a leapfrog integration method, where the $\vec E$ and $\vec H$ updates are offset by half a time step, yielding second order accuracy of the algorithm.  The spatial grid cell size and the time step in FDTD must be chosen such that they satisfy the Courant stability criterion \cite{taflove2005computational}.

Carrier motion in EMC gives rise to a current density, $\vec{J}$, which acts as a field source in FDTD; in turn, fields calculated by FDTD affect the motion of carriers in EMC. However, grid-based methods such as FDTD do not account for fields on the length scales shorter than a grid-cell size \cite{Fischetti:2001kx}, so we use the MD technique \cite{Joshi:1991vn,rapaport2004art} to calculate the short-range, sub-grid-cell fields stemming from pair-wise Coulomb interactions among electrons, holes, and ions. Carrier--ion, direct carrier--carrier, and exchange carrier--carrier (electron--electron and hole--hole) interactions are included \cite{willis:063714,emcfdtdmdgraphenejcel}.

\begin{figure}
\includegraphics[width=3 in]{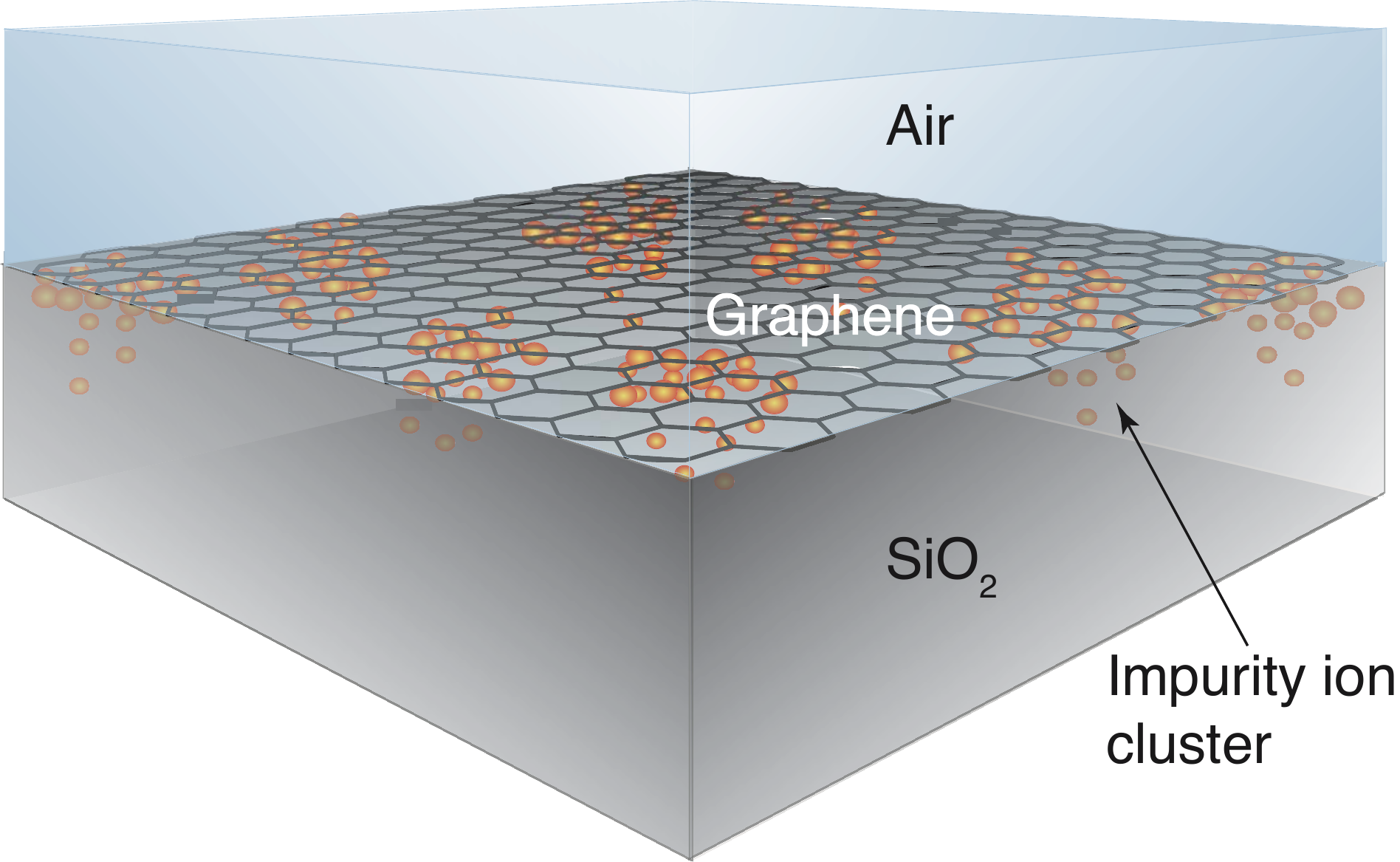}
\caption{A schematic of the simulated structure, depicting a monolayer of graphene on an SiO$_2$ substrate, with air on top. Clusters of substrate impurities near the graphene sheet are also shown.}
\label{fig_schematic}
\end{figure}

The simulated structure, shown in Figure \ref{fig_schematic}, consists of a monolayer of graphene placed on a silicon-dioxide substrate that contains charged impurities. On the four vertical planes that bound the simulation domain perpendicular to the graphene layer, we apply periodic boundary conditions to the fields and carrier momenta. The top and bottom planes that bound the simulation domain parallel to the graphene layer are terminated using convolutional perfectly matched layer (CPML) absorbing boundary conditions \cite{taflove2005computational}. In the FDTD/MD electrodynamic solver, the monolayer of graphene is defined by one plane of grid points with a dielectric constant of $2.45$, while the grid points above and below this plane are given dielectric constants of $1$ (air) and $3.9$ (SiO$_2$), respectively.

We assume that the Fermi level and carrier density in graphene can be modulated by a back gate, located at the bottom of the SiO$_2$ substrate. For a given Fermi level and temperature, the electron and hole densities are given by $n=n_\mathrm{i}\mathcal{F}_1(\eta)/\mathcal{F}_1(0)$ and $p=n_\mathrm{i}\mathcal{F}_1(-\eta)/\mathcal{F}_1(0)$, respectively.\cite{fang:092109}  Here, $n_\mathrm{i}=\frac{\pi}{6}\left(\frac{kT}{\hbar v_\mathrm{F}}\right)^2$, $\eta=E_\mathrm{F}/kT$, and $\mathcal{F}_{j}(\eta)$ is the Fermi integral of order $j$. $E_\mathrm{F}$ is the tunable Fermi level and $v_\mathrm{F}=10^8\usk\centi\meter\per\second$ is the Fermi velocity in graphene on SiO$_2$ \cite{KnoxPRB08}. The carrier ensemble in the 2D plane of graphene, comprising electrons and holes, is initialized by using random numbers to assign a position, momentum, charge, and spin to each carrier, taking into account the appropriate statistical probabilities. For the calculation of the grid-based charge density, carriers localized throughout the simulation domain are assigned to the grid using the cloud-in-cell method \cite{541446}. The initial electric field distribution is calculated by solving Poisson's equation using the successive-over-relaxation method \cite{Press:1989fk}. We use the tight-binding Bloch wave functions \cite{sule:053702} to calculate the electron-phonon scattering rates in graphene, accurately reproducing the rates from first-principles calculations \cite{PhysRevB.81.121412}, and to compute the electron-surface-optical (SO) phonon scattering rates \cite{PhysRevB.82.115452}. The scattering rates for holes are assumed to be the same as those for electrons. These initialization steps are followed by a time-stepping loop in which EMC and FDTD/MD source each other and which terminates once a steady state is achieved, as identified by the saturation of the ensemble-averaged carrier velocity and energy.

\subsection{Generating Clustered Impurity Distributions in the Simulation}\label{sec:clustered imps}

\begin{figure}
\includegraphics[width=3.3in]{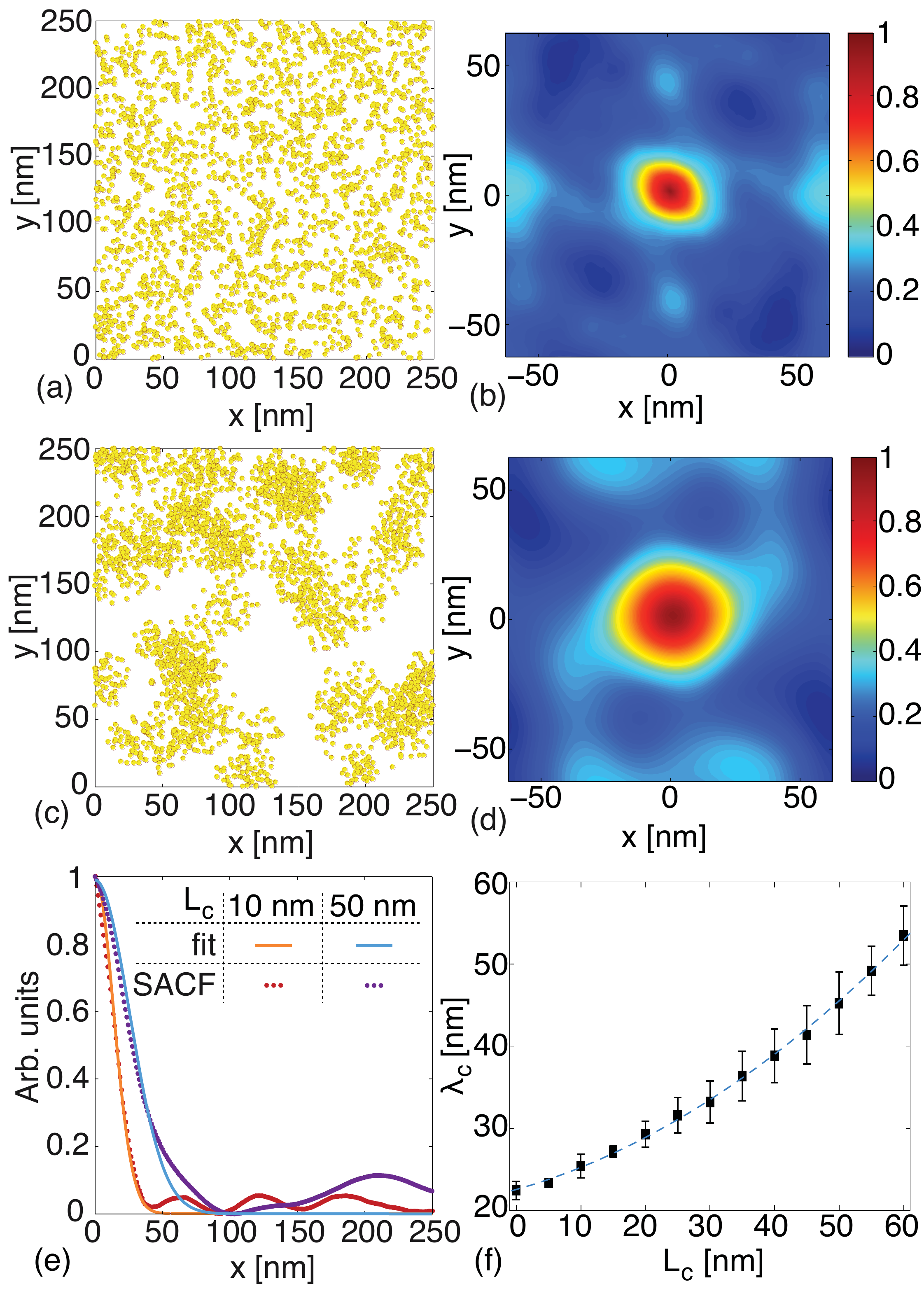}
\caption{Examples of clustered impurity distributions generated for clustering  parameters (a) $L_{\mathrm{c}}=10\usk\nano\meter$ and (c) $L_{\mathrm{c}}=50 \usk\nano\meter$. The corresponding normalized SACF are shown in (b) and (d), respectively.  The average impurity cluster size, $\lambda_{\mathrm{c}}$, is estimated from the FWHM (yellow ring) of the SACF. (e) Gaussian fits (orange and blue curves) to the SACFs from (b) and (d) (red and purple dots, respectively). (f) $\lambda_{\mathrm{c}}$ versus $L_{\mathrm{c}}$.
Each data point corresponds to the average of $14$ simulation runs for a given $L_{\mathrm{c}}$, while the error bars denote the standard deviations. The dashed line is a quadratic fit to guide the eye ($\lambda_{\mathrm{c}}=
0.005L_{\mathrm{c}}^2 + 0.22L_{\mathrm{c}} + 22.5$).}
\label{fig_impdistribution}
\end{figure}

In order to capture the influence of charged impurities on electron and hole transport in graphene, we generate different impurity distributions  throughout the SiO$_2$ substrate. The type and charge of relevant impurities vary with the processing details \cite{doi:10.1021/nn203377t}; for simplicity, we use generic impurity ions with unit positive charge. The impurity ions in the simulation are distributed in three dimensions (3D); however, impurities in the graphene literature are typically described via a cumulative sheet density, $N_{\mathrm{I}}$, in units of $\centi\meter^{-2}$. For a generated 3D distribution of ions, the sheet density is obtained by integrating over a depth equal to $2r_d$, where $r_d$ represents the effective size of an impurity ion in the MD calculation (see \citeauthor{willis:063714} \cite{willis:063714,Willis:2013uq} for more details),   followed by averaging over the total depth of the 3D distribution. $r_d$ is typically between $0.4$ and $0.8\usk\nano\meter$. We have observed that charged impurities placed deeper than $10\usk\nano\meter$ do not significantly affect carrier transport for reasonable impurity sheet densities ($N_{\mathrm{I}}<10^{12}\usk\centi\meter^{-2}$).

The problem of positioning individual impurities in 3D to achieve a predetermined cluster size distribution is related to 3D Voronoi tessellation  \cite{PrioloPRB92_Voronoi1,FanCMS04_Voronoi2,FerencPhysicaA07_Voronoi3}. Here, we have developed a relatively simple algorithm that enables us to generate an approximately Gaussian distribution of individual impurities starting from a single numerical parameter, $L_{\mathrm{c}}$, which we refer to as the \textit{clustering parameter}. For $L_\mathrm{c} = 0$, we distribute all the impurity ions stochastically according to a uniform random distribution. For a non-zero $L_\mathrm{c} $, we generate $N_\mathrm{c} = A/L_\mathrm{c}^2$ impurity clusters, where $A$ is the 2D area of the graphene layer in the simulation. To initialize the positions of individual impurities, we first distribute the centers of the $N_\mathrm{c}$ clusters stochastically. Secondly, we pick the characteristic size of each individual cluster from a uniform random distribution between $L_\mathrm{c}/3$ and $2L_\mathrm{c}/3$, the average being $L_\mathrm{c}/2$.  Next, we distribute individual impurity ions around each cluster center following a Gaussian distribution whose standard deviation equals half of the cluster size. Examples of clustered impurity distributions are shown in
Figure \ref{fig_impdistribution}a ($L_\mathrm{c}=10\usk\nano\meter$) and
Figure \ref{fig_impdistribution}c ($L_\mathrm{c}=50\usk\nano\meter$), with the corresponding
spatial autocorrelation functions (SACFs) depicted
in Figures \ref{fig_impdistribution}b and \ref{fig_impdistribution}d, respectively.
As shown in Figure \ref{fig_impdistribution}e, normalized Gaussians (orange and blue curves) fit the SACFs (red and purple dots) well. Moreover, the full-width at half-maximum (FWHM) of the SACF agrees well with the correlation length extracted from the Gaussian fits.
Henceforth, the FWHM of the impurity-distribution SACF
will be referred to as the \textit{average impurity cluster size} and denoted by $\lambda_{\mathrm{c}}$. Figure \ref{fig_impdistribution}f presents $\lambda_c$ versus $L_c$. Each data point in Figure \ref{fig_impdistribution}f represents the average of fourteen slightly different impurity ion configurations obtained stochastically for a given value of $L_{\mathrm{c}}$ (ranging from 0 to $60\usk\nano\meter$ in the increments of $5\usk\nano\meter$) and the error bars on the data points denote the standard deviations.

It is important to note that $\lambda_{\mathrm{c}}$ is conceptually different from the  correlation length $r_0$ used by \citeauthor{PhysRevLett.107.156601} \cite{PhysRevLett.107.156601} $r_0$ represents the extent to which impurity ions can interact with one another and diffuse; as a result, a larger $r_0$ results in an impurity distribution that is more spread out than clustered. In contrast, a larger $\lambda_{\mathrm{c}}$ (stemming from a larger $L_{\mathrm{c}}$, see Figure \ref{fig_impdistribution}f) represents a more clustered distribution.

\begin{figure}
\includegraphics[width=3.3in]{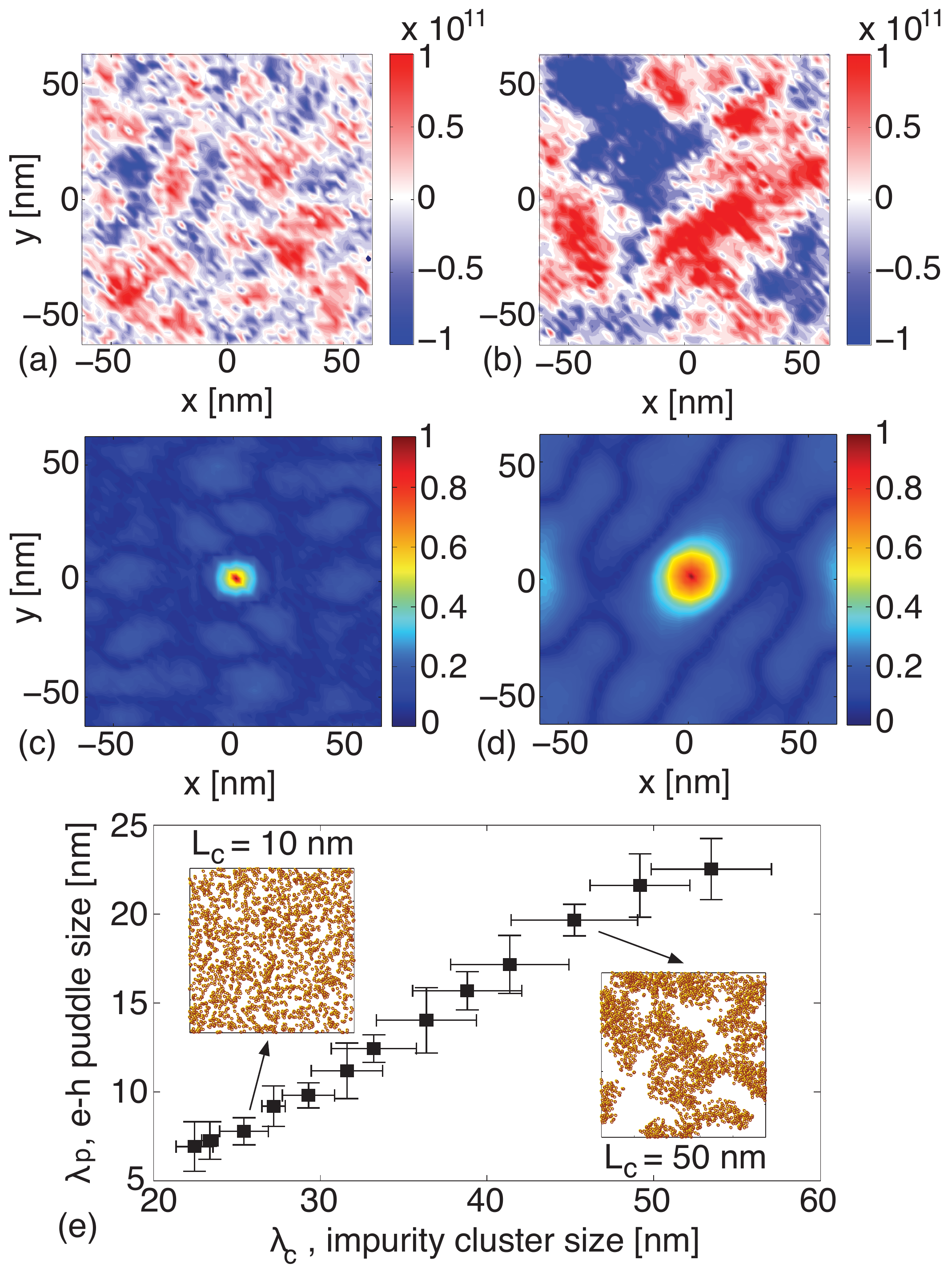}
\caption{Carrier density distribution (blue: electrons, red: holes)
depicting the electron--holes puddles formed in graphene at the Dirac point
for (a) uniform random ($L_{\mathrm{c}}=0$, $\lambda_{\mathrm{c}}=22\usk\nano\meter$) and (b) clustered
($L_{\mathrm{c}}=50\usk\nano\meter$, $\lambda_{\mathrm{c}}=46\usk\nano\meter$)  impurity distributions, both with
impurity sheet density equal to $5\times 10^{11}\usk\centi\meter^{-2}$.
The average size of the electron--hole puddles, $\lambda_{\mathrm{p}}$, is estimated from the FWHM (yellow ring) of the normalized
carrier-density SACF, shown in (c) and (d), corresponding to the random and clustered impurity distributions from (a) and (b), respectively. The estimated puddle size from (c) is $\lambda_{\mathrm{p}}=6\usk\nano\meter$ and that from (d) is $\lambda_{\mathrm{p}}=20\usk\nano\meter$.
(e) Characteristic electron--hole puddle size $\lambda_{\mathrm{p}}$
as a function of the average impurity cluster size $\lambda_{\mathrm{c}}$. Each data point corresponds to a single value of $L_{\mathrm{c}}$ (swept from 0 to $60\usk\nano\meter$ in the increments of $5\usk\nano\meter$) and is the average of $14$ simulation runs; the error bars denote the standard deviations. The insets show illustrative impurity distributions, nearly random on the left ($L_{\mathrm{c}}=10\usk\nano\meter$) and highly clustered on the right ($L_{\mathrm{c}}=50\usk\nano\meter$).}
\label{fig_puddle}
\end{figure}

\section{Results and Discussion}\label{sec:results}
\subsection{Formation of electron--hole puddles}\label{sec:e-h puddles}
Figure \ref{fig_puddle} shows the formation of electron--hole puddles in the presence of clustered impurity distributions. We simulate carrier transport at room temperature, for the Fermi level at the Dirac point ($E_{\mathrm{F}}=0$), and without external fields. The initial positions of the charge carriers in the simulation are generated  randomly based on a uniform spatial distribution and the calculated electron and hole sheet densities $n=p=8\times10^{10}\usk\centi\meter^{-2}$. As the simulation progresses, carriers move and scatter until a steady state is reached. The motion of carriers under the influence of the other charges in the domain (the clustered ions as well as other carriers) results in a charge redistribution and the formation of electron--hole puddles. The average electron--hole puddle size is estimated from the FWHM \cite{PhysRevLett.101.166803} of the SACF of the carrier density distribution. In Figures \ref{fig_puddle}a and \ref{fig_puddle}b, we contrast the carrier density distributions  that stem from the underlying uniform random ($L_{\mathrm{c}}=0$, $\lambda_{\mathrm{c}}=22\usk\nano\meter$) and clustered impurity distributions ($L_{\mathrm{c}}=50\usk\nano\meter$, $\lambda_{\mathrm{c}}=46\usk\nano\meter$). The corresponding SACFs of the carrier density are shown in Figs. \ref{fig_puddle}c and \ref{fig_puddle}d; the corresponding average electron--hole puddle sizes, estimated from the FWHM of these SACFs, are $\lambda_{\mathrm{p}}=5\usk\nano\meter$ and $20\usk\nano\meter$, respectively. These examples show a very significant difference in the sizes of electron--hole puddles that result from random and clustered impurity ion distributions. Figure  \ref{fig_puddle}e shows the average electron--hole puddle size, $\lambda_{\mathrm{p}}$, as a function of the average impurity cluster size $\lambda_{\mathrm{c}}$. Different simulation runs for the same $n$, $p$, and $N_{\mathrm{I}}$ produce slightly different puddle and impurity cluster sizes owing to the stochastic nature of the impurity position initialization and the EMC routine. Therefore, each data point in Figure \ref{fig_puddle}e represents the average of fourteen simulations for a given value of $L_{\mathrm{c}}$ (ranging from 0 to $60\usk\nano\meter$ in the increments of $5\usk\nano\meter$) and the error bars on the data points denote the standard deviations. A uniform random impurity distribution results in an average puddle size of only $6\usk\nano\meter$, while impurity clusters with an average size of 40--50 nm give rise to electron--hole puddles with an average size of $20\usk\nano\meter$, in agreement with experimental observations  \cite{PhysRevB.83.155409,Xue:2011uq,Zhang:2009qf}.

\subsection{Role of impurity distribution in carrier transport. Residual conductivity}\label{sec:imp distribution}

Next, we examine the effect of random and clustered impurity distributions on carrier transport in supported graphene. We calculate the conductivity,  $\sigma$, as a function of electron density $n$ for various spatial formations and total sheet densities of impurity ions. The electron density is varied by varying the Fermi level, mimicking the effect of a back gate. An  external \textit{dc} electric field is applied in the plane of the graphene sheet. The field is introduced using a total-field scattered-field incident-wave source condition for a uniform plane wave with a half-Gaussian temporal variation \cite{taflove2005computational}; the magnitude of the source remains constant once the peak value is achieved. The conductivity is calculated from $\sigma=\vec{J}\cdot\vec{E}/|\vec{E}|^2$, where $\vec E$  is the local electric field and $\vec{J}$ is the current density. As $\vec E$  and $\vec{J}$ are noisy, we find $\sigma$ in the steady state, upon averaging over position and time.  In the following simulation results, we have used $L_{\mathrm  c}=0$ ($\lambda_{\mathrm{c}}=22\usk\nano\meter$) for a uniform random and $L_{\mathrm{c}}=50\usk\nano\meter$ ($\lambda_{\mathrm{c}}=46\usk\nano\meter$) for a clustered impurity distribution.

\begin{figure}
\includegraphics[width=3.3in]{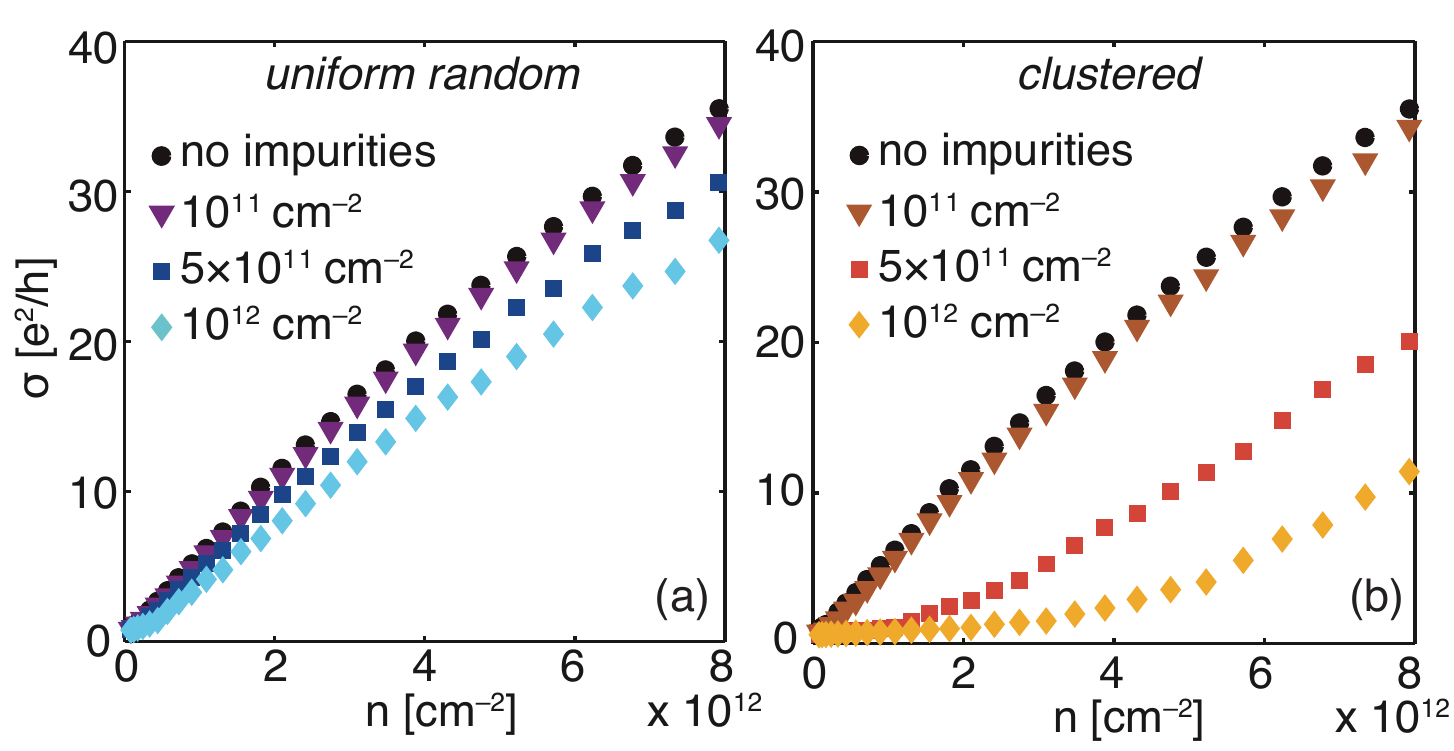}
\caption{Conductivity of graphene on SiO$_2$ for (a) uniform random ($L_{\mathrm{c}}=0$, $\lambda_{\mathrm{c}}=22\usk\nano\meter$) and (b) clustered
($L_{\mathrm{c}}=50\usk\nano\meter$, $\lambda_{\mathrm{c}}=46\usk\nano\meter$) impurity distributions,
at impurity sheet densities of $10^{11}\usk\centi\meter^{-2}$ (triangles), $5\times10^{11}\usk\centi\meter^{-2}$ (squares),
$10^{12}\usk\centi\meter^{-2}$ (diamonds), and without impurities (circles).}
\label{fig_sigmaimp}
\end{figure}

\begin{figure}
\includegraphics[width=3.3in]{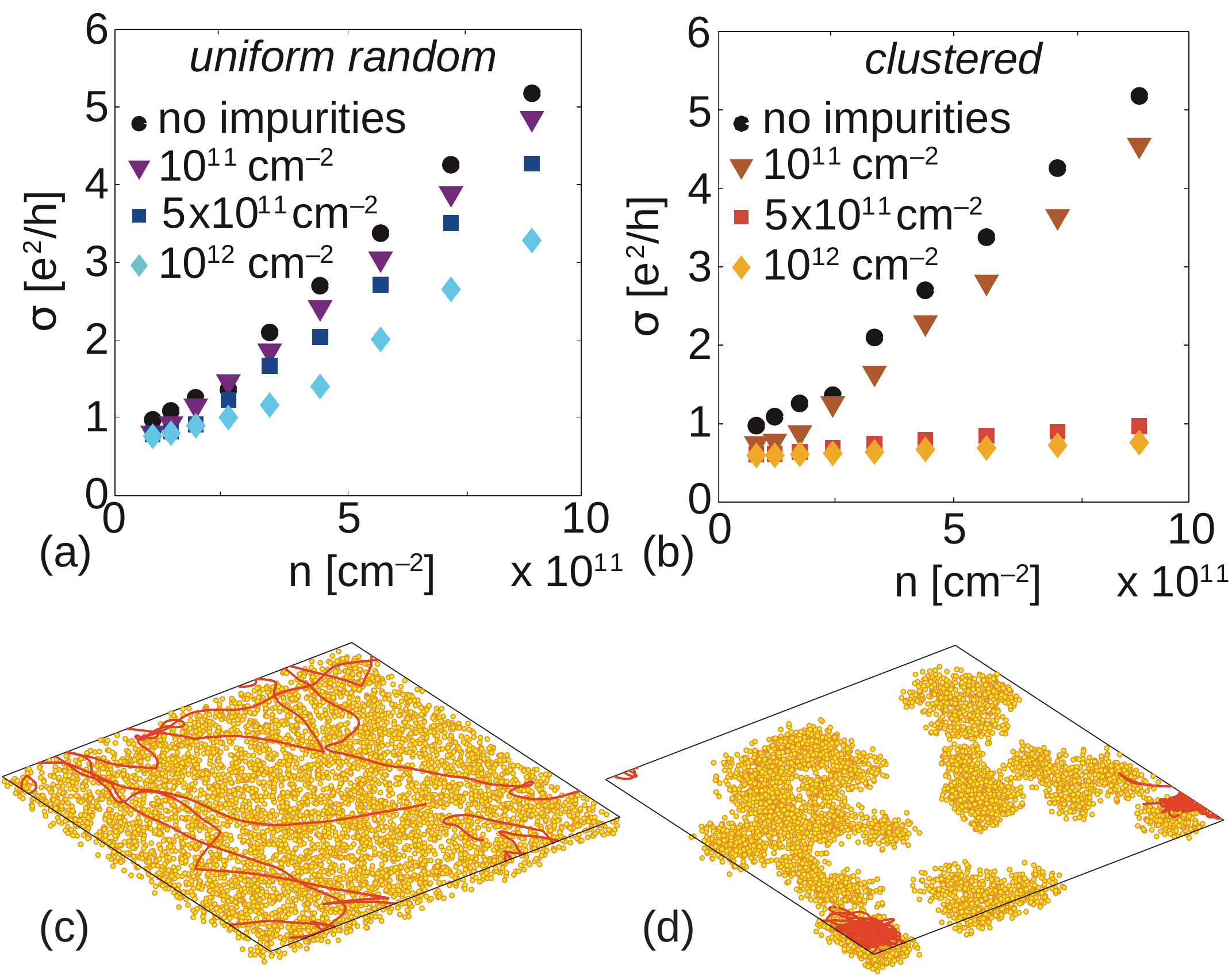}
\caption{Conductivity at low carrier densities ($<10^{11}\usk\centi\meter^{-2}$) for (a) uniform random ($L_{\mathrm{c}}=0$, $\lambda_{\mathrm{c}}=22\usk\nano\meter$) and
(b) clustered ($L_{\mathrm{c}}=50\usk\nano\meter$, $\lambda_{\mathrm{c}}=46\usk\nano\meter$) impurity distributions with sheet densities of $10^{11}\usk\centi\meter^{-2}$ (triangles), $5\times10^{11}\usk\centi\meter^{-2}$ (squares) and $10^{12}\usk\centi\meter^{-2}$ (diamonds), and no impurities (circles). Paths of sample carriers in graphene for (c) uniform random and (d) clustered impurity distributions, for $N_{\mathrm{I}}=5\times 10^{11}\usk\centi\meter^{-2}$ and $n=8.9\times 10^{11}\usk\centi\meter^{-2}$ ($E_{\mathrm{F}}=0.1\usk\electronvolt$).}
\label{fig_lownsigma}
\end{figure}

In Figure \ref{fig_sigmaimp}, we present $\sigma(n)$ for graphene on SiO$_2$  at several impurity sheet densities, ranging from impurity-free to $N_{\mathrm{I}}=10^{12}\usk\centi\meter^{-2}$, with uniform random and clustered impurity distributions. At low impurity densities ($N_{\mathrm{I}}<10^{11}\usk\centi\meter^{-2}$), the carrier-density dependence of conductivity is nearly the same for the random and clustered impurity distributions, which is not surprising and agrees with the work of Li \emph{et al.} \cite{PhysRevLett.107.156601}: with few impurities present,
their effect on transport is minor, while carrier interactions with phonons and other carriers dominate. In contrast, at impurity densities higher than $10^{11}\usk\centi\meter^{-2}$, uniform random and clustered impurity distributions result in significantly different $\sigma(n)$ variations. The most significant difference is seen at low carrier densities, where the conductivity for randomly distributed impurities increases nearly linearly with increasing carrier density, while that for clustered impurities remains flat. The slow increase in the conductivity near the charge neutrality point has also been observed in experimental measurements \cite{Chen:2008fk,2010NatNa...5..722D}, notably for samples with considerable impurity contamination.

In Figures \ref{fig_lownsigma}a and \ref{fig_lownsigma}b, we zoom in on the low-density behavior of $\sigma(n)$ from Figure \ref{fig_sigmaimp}. The low-density limit of conductivity, $\sigma_0$, known as the residual conductivity, has been observed in experiment  \cite{Chen:2008fk} and attributed to charged impurity scattering \cite{Adam20112007}. Here, we see that the value of $\sigma_0$ depends on the impurity sheet density and distribution, with higher impurity density and more clustered distributions resulting in a lower $\sigma_0$. We attribute the low-$n$ flattening of conductivity and the lower value of $\sigma_0$ for clustered distributions to carrier trapping. Figures \ref{fig_lownsigma}c and \ref{fig_lownsigma}d depict the paths of sample carriers in graphene with underlying random and clustered substrate impurity distributions, respectively. A large impurity cluster effectively traps an electron, localizing the electron's trajectory to the cluster vicinity and preventing it from participating in the current flow.


\subsection{Sublinearity in $\sigma (n)$ and carrier-carrier interactions}\label{sec:sublinearity}

\begin{figure*}
\includegraphics[width=5.5in]{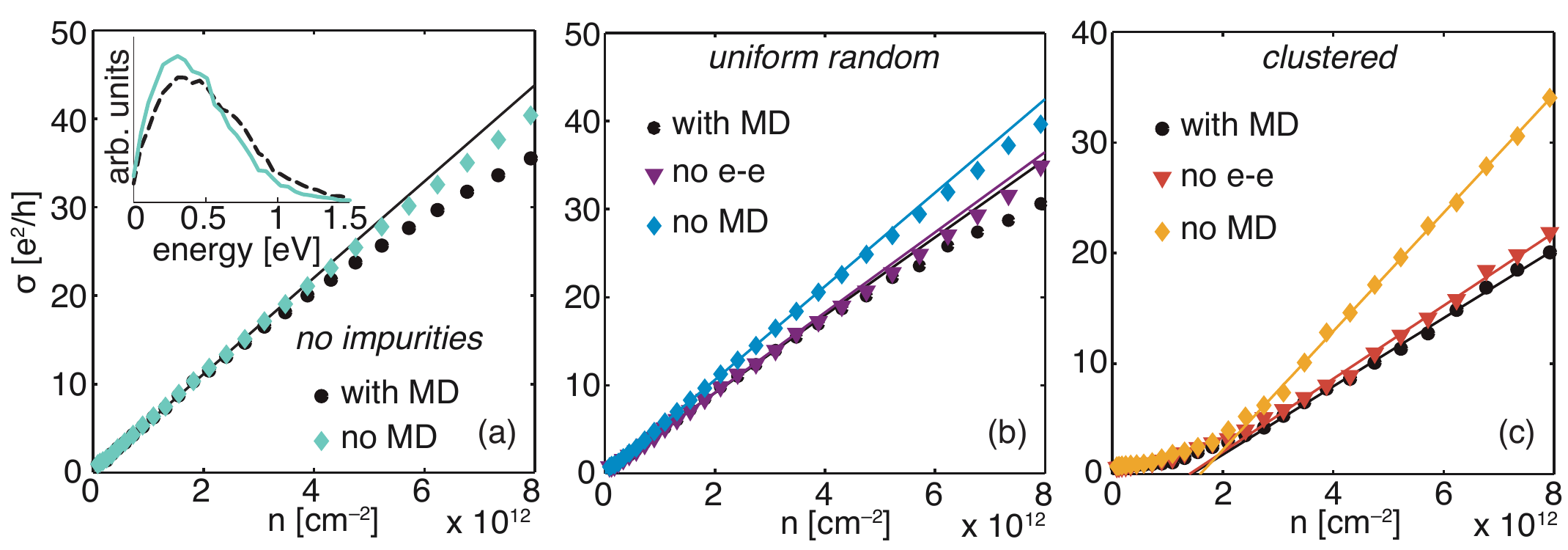}
\caption{Effect of short-range Coulomb interactions, accounted for via MD in the simulation, on transport in supported graphene (a) without substrate impurities, as well as with (b) uniform random ($L_{\mathrm{c}}=0$, $\lambda_{\mathrm{c}}=22\usk\nano\meter$) and (c) clustered ($L_{\mathrm{c}}=50\usk\nano\meter$, $\lambda_{\mathrm{c}}=46\usk\nano\meter$) impurity distributions. In all three panels, ``no MD'' indicates simulation results without any short-range interactions. In panel (a), ``with MD'' denotes simulations with the short-range direct and exchange carrier-carrier interactions included via MD. Inset to (a) shows a representative kinetic energy distribution of electrons with and without carrier-carrier interaction ($n=7\times10^{12}\usk\centi\meter^{-2}$, $E_{\mathrm{F}}= 0.3\usk\electronvolt$). In (b) and (c), ``no e-e'' indicates results of simulations with short-range carrier-ion interaction but without short-range carrier-carrier interactions, while ``with MD'' indicates simulations with the full account of all short-range interactions through the coupled EMC/FDTD/MD simulation. Impurity sheet density in panels (b) and (c) is $5\times 10^{11}\usk\centi\meter^{-2}$. }
\label{fig_sigmaMDcmp}
\end{figure*}
In Figure \ref{fig_sigmaMDcmp}, we examine the role of short-range Coulomb interactions
(carrier-carrier and carrier-ion) on \emph{dc} transport in graphene on SiO$_2$. We account for these effects via the MD part of the simulation and can selectively turn them on or off to better elucidate their importance.
Figure \ref{fig_sigmaMDcmp}a presents $\sigma(n)$ for impurity-free graphene, with MD (circles) and without MD (diamonds); without impurities, MD accounts only for the short-range, direct and exchange carrier-carrier interactions. We deduce that the sublinearity in $\sigma(n)$ at high carrier densities occurs
largely due to carrier-carrier interactions: when we exclude their short-range component by turning off MD, $\sigma(n)$ becomes nearly linear. Any remaining sublinearity in the ``no MD'' results can be attributed to the long-range,  direct carrier-carrier Coulomb interaction that is captured by the FDTD solver. Carrier-carrier Coulomb interactions do not directly affect conduction (the total momentum of an interacting pair is conserved, as is the pair's total energy), but redistribute the momentum and energy among the pair and therefore affect the single-particle distribution function, pushing it towards a shifted Fermi-Dirac distribution  \cite{KrimanPRL90,LugliPRL86,LugliTED85,Lundstrom}.
The inset to Figure \ref{fig_sigmaMDcmp}a presents the computed distribution of electrons over kinetic energy with and without carrier-carrier interaction for the electron density of $7\times10^{12}\usk\centi\meter^{-2}$ ($E_{\mathrm{F}}= 0.3\usk\electronvolt$). This curve corresponds to $g(E)f(E)$, where $g(E)$ is the electron density of states and $f(E)$ is the distribution function, and carrier-carrier interaction clearly leads to a greater abundance of higher-energy carriers. Since electron and hole scattering rates with phonons increase with increasing energy, the redistribution of carriers over energy effectively raises the average carrier-phonon scattering rate and leads to a reduction in conductivity that we observe as the slopeover in $\sigma(n)$.

In Figures \ref{fig_sigmaMDcmp}b and \ref{fig_sigmaMDcmp}c, we plot  $\sigma(n)$ for uniform random and clustered impurity distributions with all short-range interactions accounted for through MD (circles, ``with MD''), with short-range carrier-ion but without carrier-carrier interactions (triangles, ``no e-e''), and without any short-range interactions (diamonds, ``no MD''). We have already discussed the low-\textit{n} region (see Figure \ref{fig_lownsigma}) and will focus here on the medium-to-high electron density range. In both Figures \ref{fig_sigmaMDcmp}b and \ref{fig_sigmaMDcmp}c, the sheet density of impurities is appreciable ($5\times 10^{11}\usk\centi\meter^{-2}$), so carrier-ion interactions govern transport in the medium and the $\sigma(n)$ dependence is largely linear \cite{tsuneya2006screening}. Turning off short-range carrier-carrier interactions causes insignificant change to the slope in either panel, while turning off short-range carrier-ion interactions significantly affects the slope.

\subsection{Estimating impurity density from the inverse slope of $\sigma (n)$}\label{sec:inverse slope}

\begin{figure}
\includegraphics[width=3.0in]{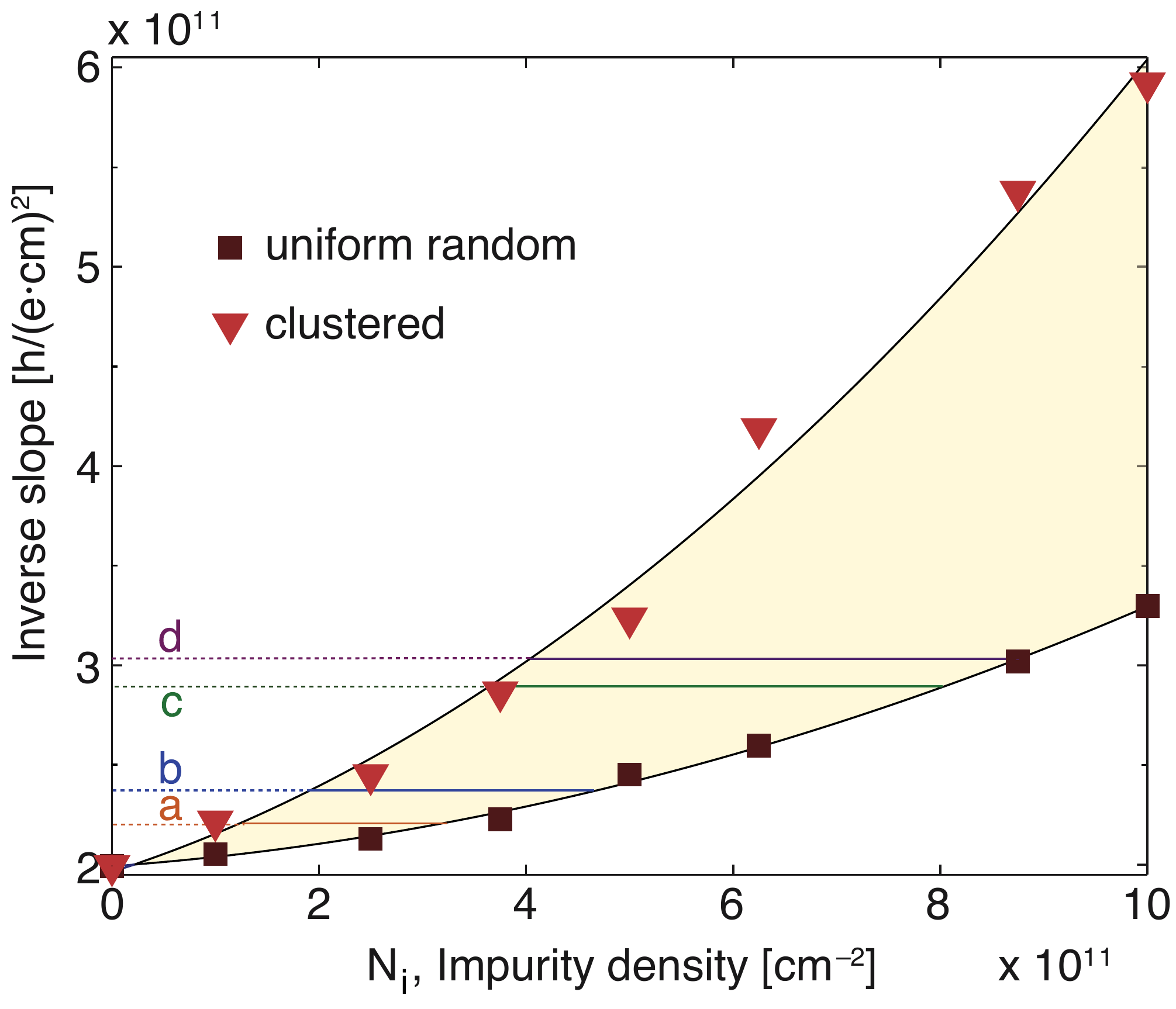}
\caption{Inverse slope of $\sigma(n)$ as a function of the sheet impurity density for graphene on SiO$_2$ at room temperature. Squares denote the uniform random impurity distribution ($L_{\mathrm{c}}=0$, $\lambda_{\mathrm{c}}=22\usk\nano\meter$), while triangles correspond to clustered impurity distributions that would give realistic e-h puddle sizes ($L_{\mathrm{c}}=50\usk\nano\meter$, $\lambda_{\mathrm{c}}=46\usk\nano\meter$). The horizontal lines (a--d) correspond to the inverse slope values obtained in several experiments: a -- Ref. \cite{VicarelliNatMat2012}, b -- Ref. \cite{Schedin:2007zr}, c -- Ref. \cite{NovoselovPNAS2005},  and d -- Ref. \cite{NovoselovScience2004}. The $N_i$-range between the intercepts of an inverse-slope horizontal line with the clustered and random distribution curves  (i.e. the range within the lightly shaded area) yields an estimate of the impurity density range.}
\label{fig_impestimate}
\end{figure}

The slope of the $\sigma(n)$ curves in the linear region is governed by the short-range carrier-ion interactions, and is dependent on both the impurity density (Figure \ref{fig_sigmaimp}) and distribution (Figures \ref{fig_sigmaMDcmp}b and \ref{fig_sigmaMDcmp}c). As the slope can be accurately measured in experiment, we can use it to indirectly extract the impurity density and cluster size. In Figure \ref{fig_impestimate}, we present the EMC/FDTD/MD simulation results for the inverse slope of $\sigma(n)$ in the linear region as a function of the sheet impurity density, with the cluster size as a parameter. The solid markers represent the simulation results for uniform random ($L_{\mathrm{c}}=0$, $\lambda_{\mathrm{c}}=22\usk\nano\meter$, denoted by squares) and clustered impurity distributions ($L_{\mathrm{c}}=50\usk\nano\meter$, $\lambda_{\mathrm{c}}=46\usk\nano\meter$, denoted by triangles). The curves are polynomial fits to guide the eye and indicate the range of results for different impurity distributions. As we discussed earlier, impurity cluster sizes of 40--50 nm correspond to electron--hole puddle sizes obtained in experiment(see Fig. \ref{fig_puddle}), so it is likely that a reasonable sheet impurity density estimate can be obtained from the clustered impurity curve in Figure \ref{fig_impestimate}.
As examples, we present the inverse slopes extracted from several room-temperature measurements on supported graphene (a -- Ref. \cite{VicarelliNatMat2012}, b -- Ref. \cite{Schedin:2007zr}, c -- Ref. \cite{NovoselovPNAS2005},  and d -- Ref. \cite{NovoselovScience2004}). The intercepts of each inverse-slope horizontal line with the clustered and random distribution curves in Fig. \ref{fig_impestimate} indicate an estimate of the impurity-density range, with the clustered-curve intercept likely yielding a good approximate value for $N_i$.  Note that the more recent experiments  (a from 2012  \cite{VicarelliNatMat2012} and b from 2007 \cite{Schedin:2007zr}), which arguably had samples with fewer impurities than the early ones owing to the advances in processing, indeed correspond to lower sheet impurity densities than the earlier measurements (c in 2005 \cite{NovoselovPNAS2005} and d in 2004 \cite{NovoselovScience2004}).

\section{CONCLUSION}\label{sec:conclusion}

In summary, we have employed EMC/FDTD/MD coupled simulation of carrier  transport and electrodynamics to investigate the effects of carrier-carrier and carrier-ion Coulomb interactions on the transport properties of graphene on SiO$_2$, with focus on the role of substrate impurity clustering. While corrections due to many-particle correlations \cite{Martin:2008ve} and coherent transport features \cite{young2009quantum,MuccioloJPCM2010} may play an important role in extremely clean suspended graphene at low temperatures, our simulations accurately capture the physics of diffusive, room-temperature carrier transport in supported graphene, which is relevant for device applications. We have shown that clustered impurity distributions with an average cluster size of 40--50 nm result in the formation of electron--hole puddles with a typical size of $20\usk\nano\meter$, comparable to observed values. We have also demonstrated that high-density clustered impurities lead to carrier trapping and a flattening of the low-$n$ $\sigma(n)$  dependence.
By selectively controlling the short-range Coulomb interactions of the carriers in the coupled EMC/FDTD/MD simulation, we have shown that the sublinear $\sigma(n)$ dependence at high carrier densities can be attributed to carrier-carrier interactions  \cite{Chen:2008fk,2010NatNa...5..722D}.
The slope of the linear-region $\sigma(n)$ relates to the strength of the carrier-ion Coulomb interactions, and we have characterized its dependence on the impurity density and distribution. The computed dependence of the linear-region slope of $\sigma(n)$ on the impurity density might be used as a noninvasive technique for estimating the impurity density in experiment.

\section*{Acknowledgment}
This work has been primarily supported by AFOSR, award No. FA9550-11-1-0299. I.K. acknowledges partial support by NSF, award No. 1201311.



%

\end{document}